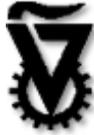

# Technion – Israel Institute of Technology

## Electrical Engineering Department
Communication Laboratory

# GNSS Differential Interferometer

**December 2010**

<u>Supervisors</u>:   Mr. Avner Kaidar
              Dr. Gilad Even-Tzur

<u>Students</u>:      Igor Slutsker
              Boris Katsman

# Contents

# 1  Abstract

The Global Positioning System (GPS) is a global navigation satellite system that provides reliable location and time information on or near the Earth, given an unobstructed line of sight to four or more GPS satellites. It is composed of 24 to 32 Satellite Vehicles (SVs) in medium Earth orbit (~22,000 Km above earth). The GPS network was established in 1973 by the United States government and is freely accessible by anyone with a GPS receiver.

Baseline azimuth calculation using GPS carrier phase data can be achieved in a classic way by using two or more geodetic synchronized GPS receivers. Such system includes dedicated expensive components and thus is not applicable to many uses.

The development of low cost GPS receivers have come to a point where good quality carrier phase measurements are available and double differencing systems utilizing these receivers have proven feasible. Inexpensive receivers like this, combined with attitude determination algorithm, can make accurate attitude determination using GPS accessible to everyone.

Our project consisted of developing a low cost attitude determination system utilizing 2 Fastrax off-the-shelf receivers. Phase and code data from the receivers is processed, filtered and used for attitude determination using double-difference algorithm.

Possible applications for such systems are for example:
- Civilian – Search and rescue gear
- Industrial – Low cost geodetic measurement
- Military – Inexpensive weapon guidance

# 2  Project Goal

The goal of the project was to design and manufacture a low cost offline GPS based differential compass capable of measuring accurate directional vector baseline between distinctive GPS receivers.

# 3  Introduction

In order to build a GPS differential compass, we've chosen Fastrax© IT03 GPS receiver based on the RF chip uN8021 and the base-band Atheros© uN8130 chip. This receiver was chosen for its ability to store carrier phase measurements and its low cost.

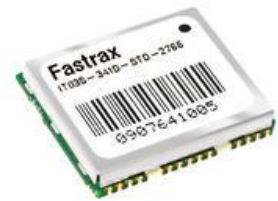

Our Project was composed of 2 major parts:

1.  Design  and production of an IT03 receiver board:
    a.  Theoretical study and understanding of receiver peripheral requirements.
    b.  Board design, PCB layout.
    c.  Physical assembly of elements after production.
    d.  Integration with SDK software
    e.  Testing of the board.

2.  MATLAB algorithms:
    a.  Offline conversion of the recorded iTalk protocol binary files into RINEX protocol files.
    b.  Interfacing  the RINEX files into Double Differential calculation algorithm [6].
    c.  Field tests
    d.  Analysis of results.

# 4  Theoretical Background

## 4.1 Length of propagation path

Let **k**,**m** be the locations of two GPS receiver antennas, **b** – baseline vector between them. The length of the propagation paths between SV **p** and receiver **k**, or SV **p** and receiver **m** in terms of fractional and integer carrier cycles are:

$$\Phi_k^p(t) = \phi_k^p(t) - \phi^p(t) + N_k^p + S_k^p + f\tau_p + f\tau_k - \beta_{iono} + \delta_{trop}$$

$$\Phi_m^p(t) = \phi_m^p(t) - \phi^p(t) + N_m^p + S_m^p + f\tau_p + f\tau_m - \beta_{iono} + \delta_{trop}$$

Where:

- **k** and **m** refer to the receiver antennas phase centers.
- **p** is the satellite vehicle (SV) signal source.
- $\phi^p(t)$ is the transmitted satellite signal phase as a function of time.
- $\Phi_k^p(t)$ and $\Phi_m^p(t)$ are the receiver-measured satellite signal phase as a function of time.
- N is the unknown integer number of carrier cycles from SV **p** to **k** or SV **p** to **m**.
- **S** is the phase noise due to all noise sources (e.g receiver, multipath…)
- **f** is the carrier frequency
- $\tau$ is the associated satellite or receiver clock bias
- $\beta_{iono}$ is the advance of  the carrier (cycles) due to the ionosphere.
- $\delta_{tropo}$ is the advance of the carrier (cycles) due to the troposphere.

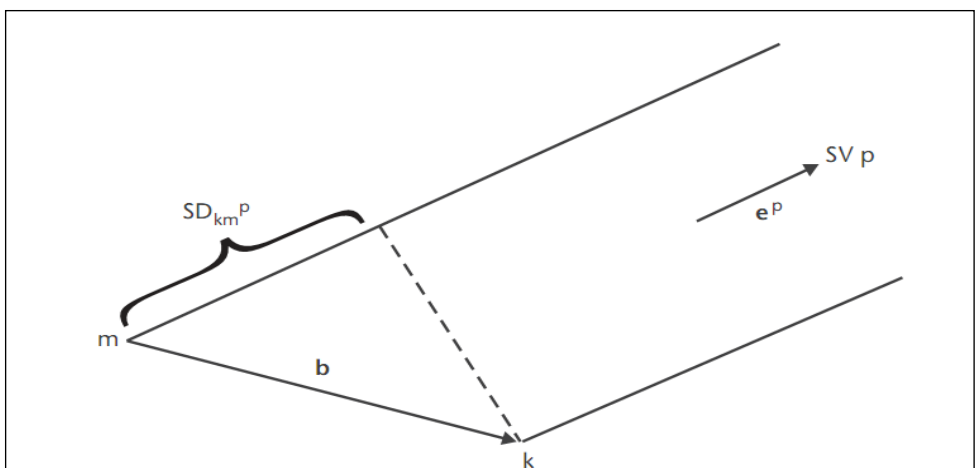

**Figure 1 – propagation path from SV p**

## 4.2 Single difference

By differencing the propagation paths from receivers **k** and **m** to SV **p** we create the **S**ingle **D**ifference (SD) equation for SV **p**:

$$SD_{km}^p = \phi_{km}^p(t) + N_{km}^p + S_{km}^p + f\tau_{km}$$

And by differencing the propagation path from receivers **k** and **m** to SV **q** we create the SD equation for SV **q**:

$$SD_{km}^q = \phi_{km}^q(t) + N_{km}^q + S_{km}^q + f\tau_{km}$$

Using SD equations we are able to:

1) Cancel the effect of SV clock phase bias
2) Merge Receiver phase biases to a single bias $\phi_{km}^q(t)$
3) Cancel the delay effect through Troposphere and Ionosphere

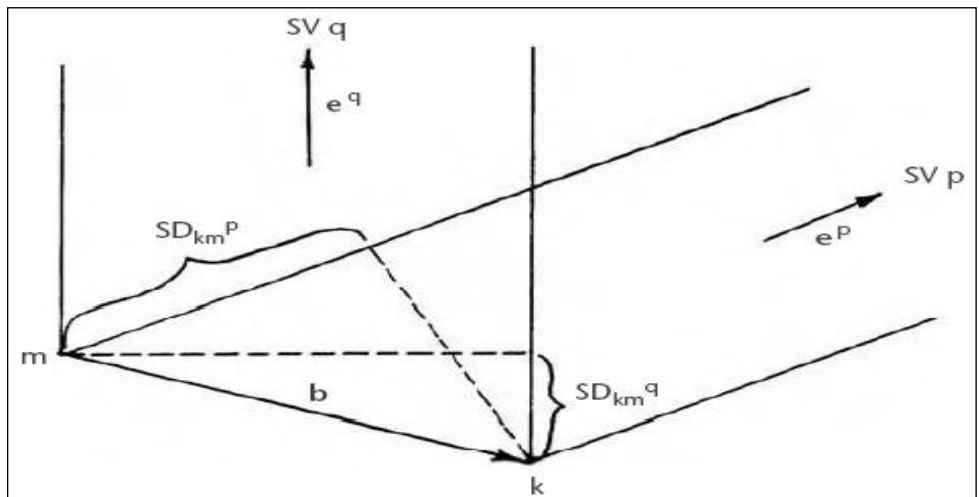

**Figure 2 – Single Difference**

The projection of **b** on the **L**ine **O**f **S**ight (LOS) between **p** and **m** can also be written as the inner (dot) product of **b** with a unit vector in the direction of SV **p** (same for SV **q**), and by converting to wavelengths we get the following SD equations:

$$SD_{km}^p = (b \bullet e^p)\lambda^{-1} = \phi_{km}^p(t) + N_{km}^p + S_{km}^p + f\tau_{km}$$
$$SD_{km}^q = (b \bullet e^q)\lambda^{-1} = \phi_{km}^q(t) + N_{km}^q + S_{km}^q + f\tau_{km}$$

## *4.3 Double difference*

Now we difference the two SD equations for SV **p** and SV **q** and we get the **D**ouble **D**ifference (DD):

$$DD_{km}^{pq} = (b \bullet e^{pq})\lambda^{-1} = \phi_{km}^{pq} + N_{km}^{pk} + S_{km}^{pk}$$

Using the DD calculation we cancel out the receivers clock bias, since it can be assumed that it remains (almost) the same while measuring for both SVs. Forming the same set of equations for 5 different SVs we can fill in the following matrix:

$$\begin{bmatrix} DD_{cp1} \\ DD_{cp2} \\ DD_{cp3} \\ DD_{cp4} \end{bmatrix} = \begin{bmatrix} e_{12x} & e_{12y} & e_{12z} \\ e_{13x} & e_{13y} & e_{13z} \\ e_{14x} & e_{14y} & e_{14z} \\ e_{15x} & e_{15y} & e_{15z} \end{bmatrix} \begin{bmatrix} b_x \\ b_y \\ b_z \end{bmatrix} + \begin{bmatrix} N_1 \\ N_2 \\ N_3 \\ N_4 \end{bmatrix} \lambda$$

Where (1st line explained):
1) $DD_{cp1}$ represents the first DD measurement
2) $e_{12}$ represents the differenced unit vector between the two satellites under consideration (here – SV1, SV2).
3) b is the baseline vector we are searching.
4) N is the associated integer carrier cycles ambiguity.
5) The wavelength λ is used to provide consistency with $DD_{cp1}$ and b that are here in meters.

## *4.4 System Description*

The project setup is composed of 2 identical GPS receiver boards, connected to USB ports, one is defined as "Master" and the other as a "Slave". The data from the boards is sent in iTalk[5] protocol to a PC. The iTalk messages are archived into .itk binary files. The data is then processed in 2 steps:

The first step is converting the raw data files from iTalk to RINEX[4] format. This process is initiated with a user-friendly GUI (where the user edits setup and conversion parameters). The data files are then processed - reading iTalk files, synchronizing between them and finding mutual measurement epochs of master and slave receivers.

Further we apply a search algorithm that finds the longest available sequence of stable SV (Available throughout the whole period selected for conversion),that are present in both the master and the slave receivers (at this stage we neglect the sequences where non-valid data was received).
Next we convert the chosen sequence to RINEX data files (See Appendix C for details).
This step is described in Appendix D.

The second step is determination algorithm for the baseline vector in the following order:
1. Smoothing of the code data using carrier phase data.
2. Single point positioning calculation for master antenna.
3. Differential positioning – baseline magnitude and error calculation.
4. 3D attitude determination using direct computation method.

The result of this process is the displayed azimuth value and baseline calculated error.

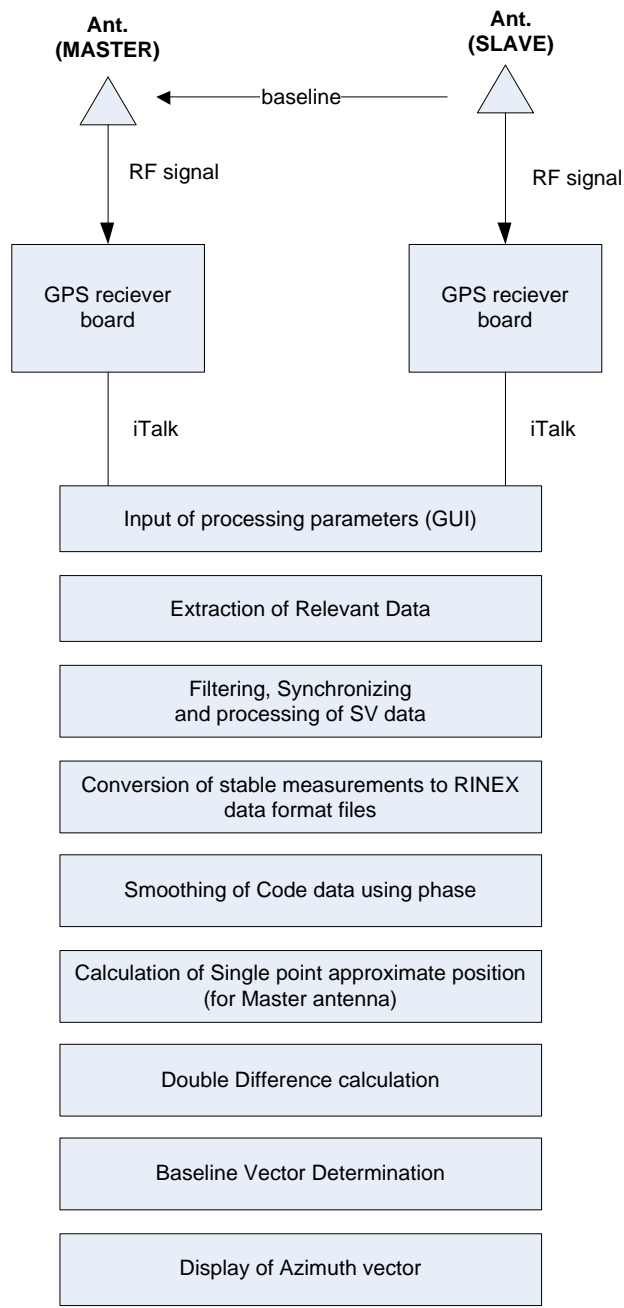

**Figure 3 -System Flow Chart**

# 5  Board Design

Each circuit board provides peripherals for a single Fastrax 03 receiver.
The peripherals include:

A. 2 ultra-stable DC-DC regulators fed from 1 or 2 USB supply lines.
B. 2 USB-UART converters.
C. ESD protection for both USB ports.
D. 1 buffer for external 1PPS drive.
E. 3 LEDs for receiver status indications.
F. 3 LEDs for Power regulators and 1PPS status.
G. SMA RF connector for GPS antenna input (Active or Passive antenna)

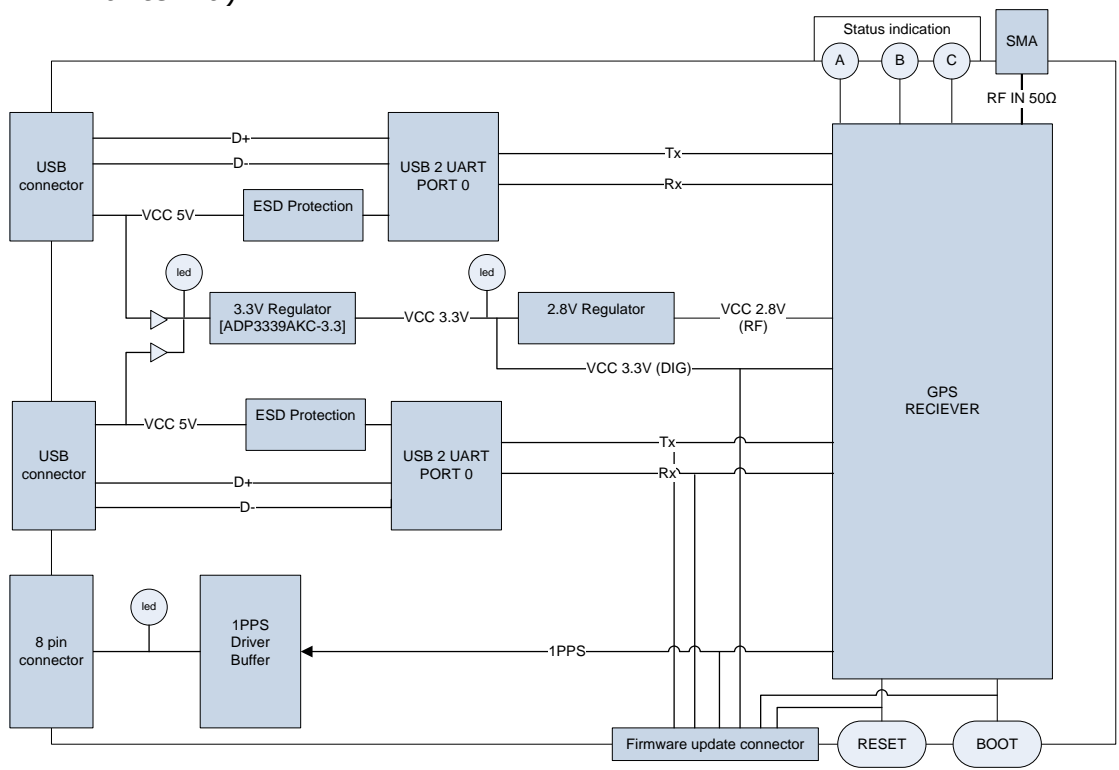

**Figure 4 - Circuit Block Diagram**

For further description of the board design, see <u>Appendix B</u>

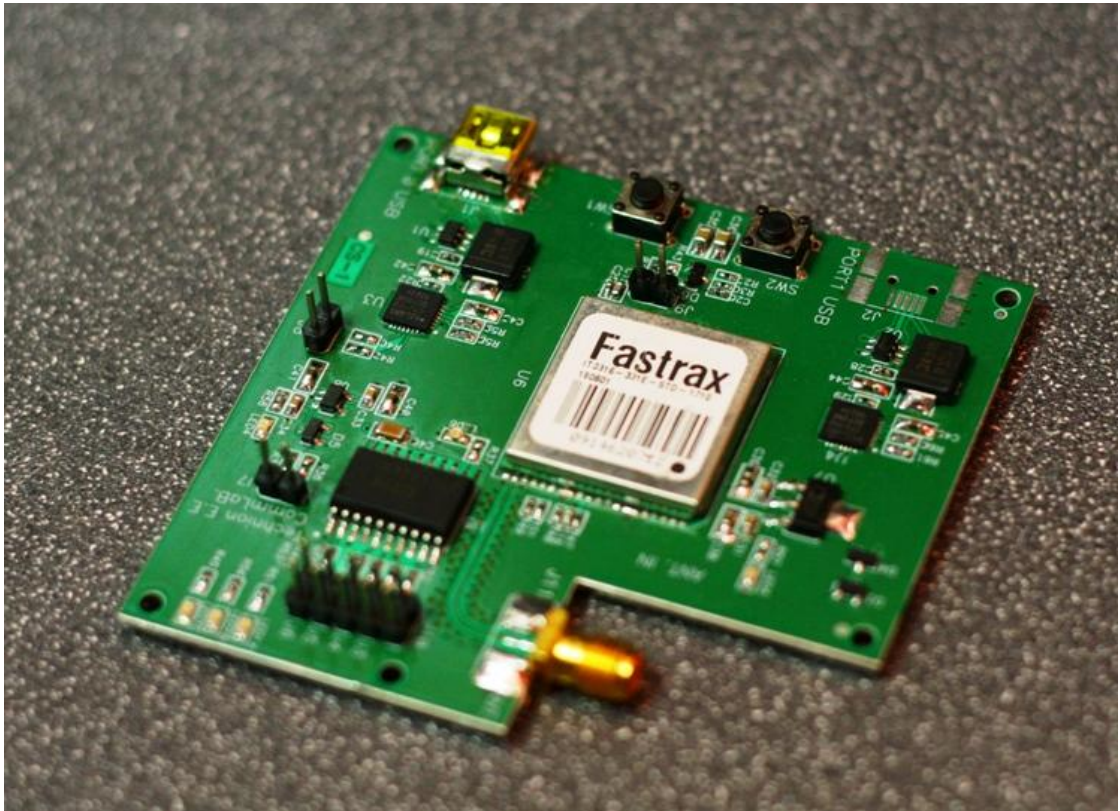

**Figure 5 - Final PCB photo**

# 6 Problems and Resolutions

## *Problems and resolutions in board design*

**Problem:** In the  first prototype board  the iTrax03-S receiver didn't fit the physical layout on the board. The purchased receiver was actually iTrax03, it was designed for  iTrax03-S. Those two receivers have different physical dimensions and pin-out:

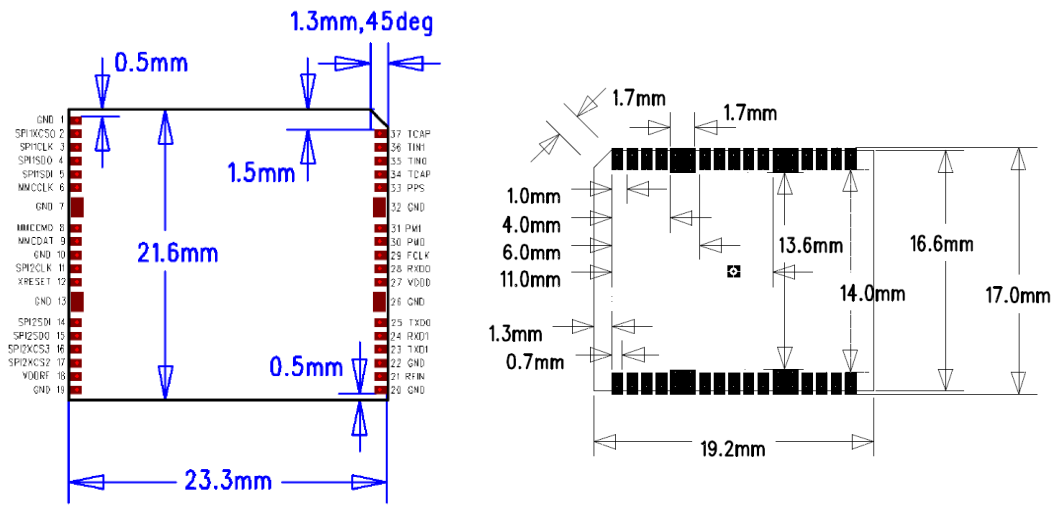

**Figure 6 – iTrax03 (left), iTrax03-S (right)**

**Solution:**
1. We've build an interface board, matching the I/O of both packages, and connected the iTrax03 receiver to the board using wire-up in order to proceed with the tests.
2. The PCB layout was  changed to fit the correct receiver version (iTrax03) and sent the new layout to production.

**Problem:** LED_B was not connected properly on the board.
**Solution:** Wired-up for testing and changed in the final layout.

## *Problems and resolutions in attitude determination*

**Problem:**

The main data used for baseline determination of the carrier phase is calculated by the following formula:

Φ(cycles) = dwCarrCount + wCarrPhase / 256 [8]

By using the above formula, we get the following results (below), that don't meet our expectations. We expect continuous rising/falling behavior of the phase. Instead, the formula generates raising chainsaw type graph, with ~400 epoch interval for every SV.

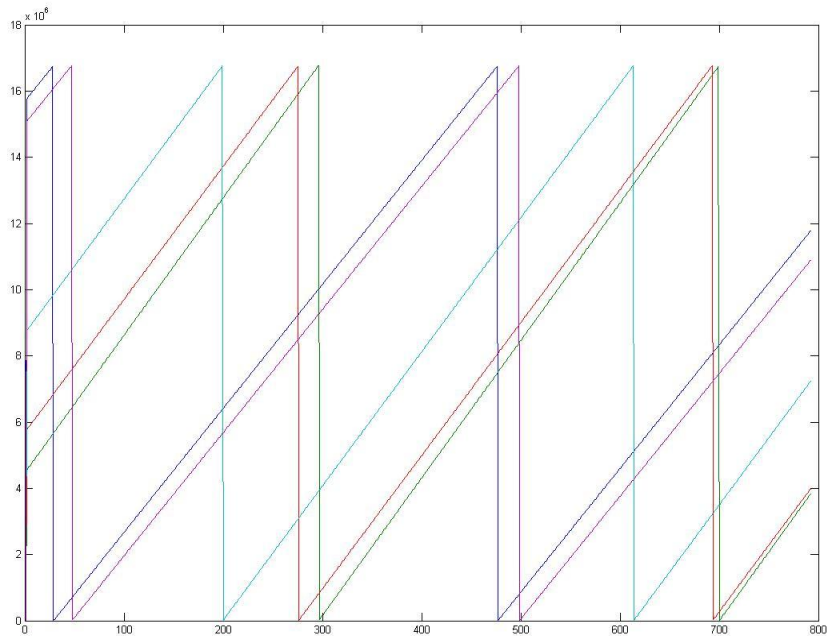

**Figure 7 - Calculated Phase − 5 SVs**

**Solution:**  After consulting our supervisors and Fastrax technical support, We found out that  that the phase processing in Fastrax receivers is performed at IF frequency of 38596.34 Hz and the time between epochs is actually 1.000648 sec (instead of 1sec.) Both issues are not documented. Therefore we need to integrate the phase over epochs and subtract between each 2 consecutive  epochs in order to get the correct result for Phase measurement.

**Problem:** Non consistent results – sometimes the algorithm converged to the correct results, and sometimes it converged to wrong results or didn't converge at all.

**Solution:** We performed a few steps in order to find the cause of the matter:

- Comparing RINEX navigation file generated by our code from received .itk files to the Geodetic accurate Ashtec receivers' navigation file for the same time and position – <u>The files were identical</u>.
- Comparing .itk pseudo-range measurements provided directly by our receiver to simulated environment measurements provided by iGalileo simulator – The files were not identical, but the <u>differences in distances were very similar </u>(up to 500 m).

**Conclusion**

- After extensive investigation of the cause for inconsistency, , we could not find a direct cause apart from an assumption that the pseudo-range value provided by the iTrax03 receiver is not a raw measurement, but rather some undocumented corrections were applied.
- This could cause a double-correction and therefore create an unstable single point positioning for the Master antenna, eventually causing a wrong attitude calculation.

# 7 Field Tests

Several field tests were performed.
All tests were performed on the same location – Carmel beach, Haifa, Israel, in order to eliminate possible location-based error factors.
Different baseline angles and different baseline length were tried in order to understand the implications of baseline length on calculation accuracy – 0.5,1,2,5,10 m.

The tests were performed using the following equipment:
2 x IT03 receiver board
2 x Laptop computer with Fastrax GPS workbench 4 software
2 x USB cable
2 x Active GPS L1 antenna with 2m cable and an SMA connector

At the end of the measurement phase the results were recorded and processed using our algorithm. RINEX files were generated and fed into the Double Difference [6] processing algorithm.
Results were compared to the mean value of the baseline vector azimuth that was recorded for each of the performed tests, using external magnetic compasses (after magnetic anomaly correction).

## *7.1 Converging Measurement #1*

Date :   2010/10/18
# of SVs :       5
Baseline :       2m
Bearing :        ~0°

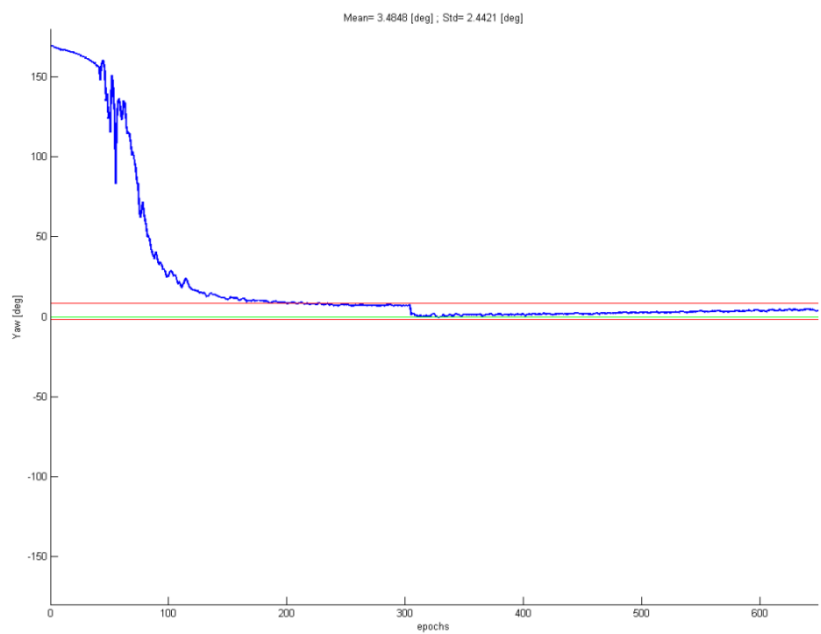

**Figure 8 - Converging Measurement**

## *7.2 Converging Measurement #2*

Date :  2010/10/18
# of SVs :     7
Baseline :     2m
Bearing :      ~45°

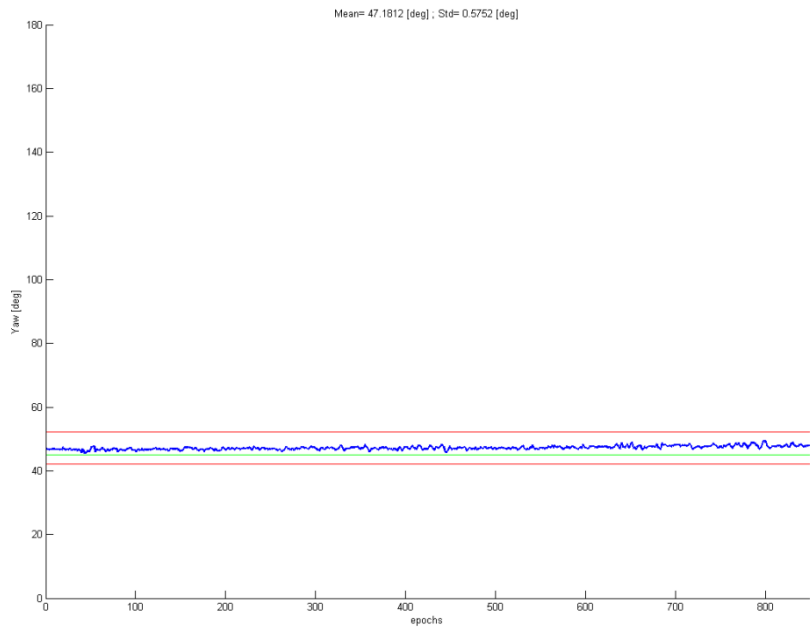

**Figure 9 - Converging Measurement #2**

## *7.3 Non-Converging Measurement*

Date :  2010/11/13
# of SVs :     6
Baseline :     2m
Bearing :      ~020°

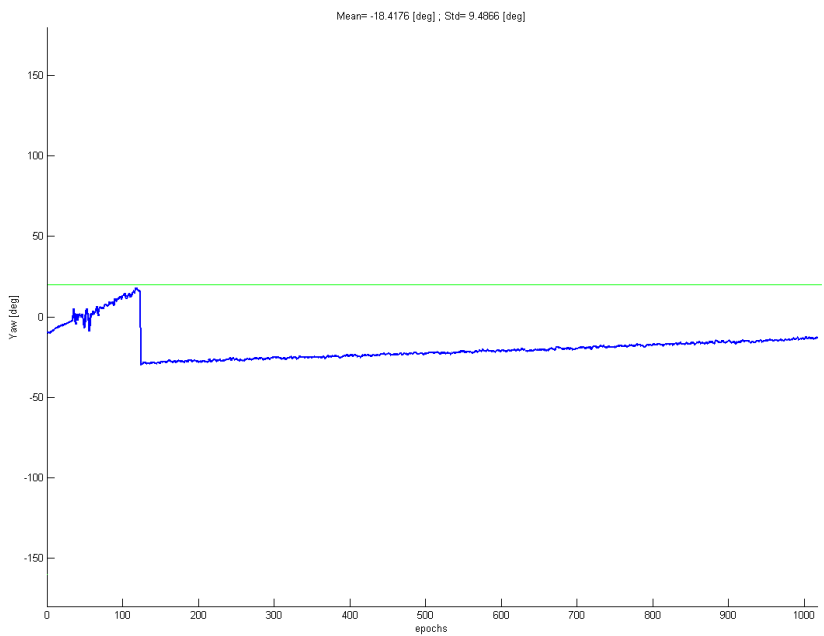

**Figure 10 - Non Converging Measurement**

## 7.4 Wrong Converging Measurement

Date :  2010/11/21
# of SVs :      6
Baseline :      5m
Bearing :       ~016°

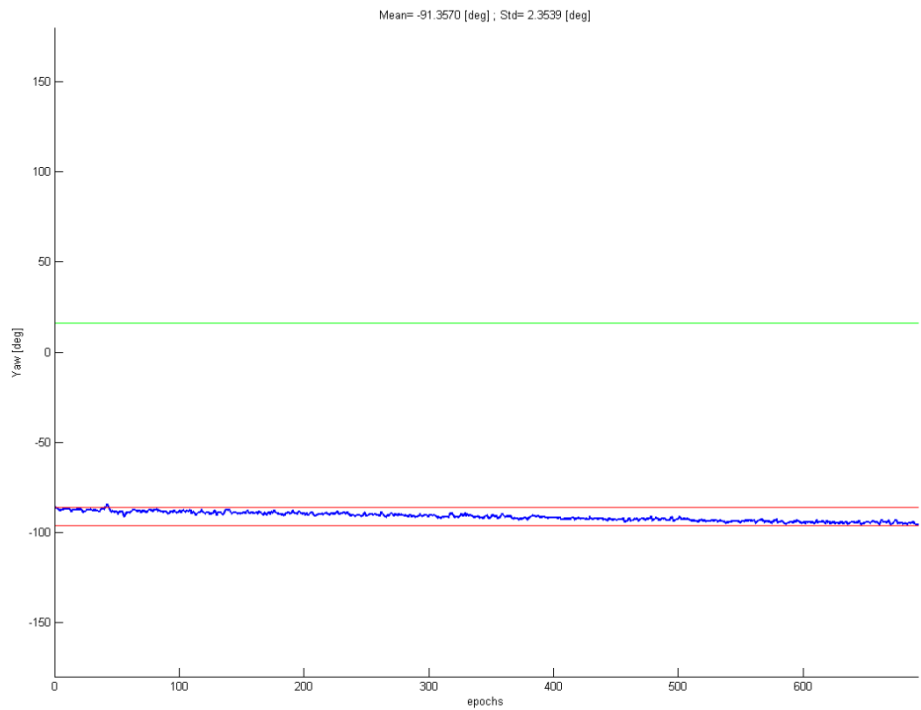

**Figure 11 - Wrong Converging Measurement**

# 8 Abbreviations

| | |
|---|---|
| BOM | Bill Of Materials |
| SD | Single Difference |
| DD | Double Difference |
| ESD | Electro Static Discharge |
| GPS | Global Positioning System |
| LED | Light Emitting Diode |
| NMEA | National Marine Electronics Association |
| PCB | Printed Circuit Board |
| PPS | Pulse Per Second |
| RINEX | Receiver Independent Exchange Format |
| RF | Radio Frequency |
| SD | Single Difference |
| SDK | Software Development Kit |
| SMA | Sub Miniature version A (connector type) |
| SV | Satellite Vehicle |
| UART | Universal Asynchronous Receiver Transmitter |
| USB | Universal Serial Bus |

# 9  References & Bibliography

# Appendix A - BOM

The following list describes the Bill Of Materials used in the GPS board:

| Item | Quantity | Reference | Part | PCB footprint |
|---|---|---|---|---|
| 1 | 1 | C15 | 18pF | CC0402(M) |
| 2 | 1 | C16 | 10nF | CC0402(M) |
| 3 | 4 | C19,C25,C26,C28 | 100nF | CC0402(M) |
| 4 | 4 | C31,C32,C33,C34 | 100nF | CC0603(M) |
| 5 | 2 | C35,C36 | 4.7uF | CC0805 |
| 6 | 8 | C38,C39,C40,C41,C42,C43,C44,C45 | 1uF | CC0805 |
| 7 | 1 | C46 | 220nF | CC0805 |
| 8 | 2 | D1,D2 | B320A | CASE D |
| 9 | 4 | D3,D4,D5,D6 | BAT54/PLP | SOT-23(1-3) |
| 10 | 2 | J1,J2 | MOLEX675031020 | |
| 11 | 8 | J3,J4,J5,J6,J7,J8,J9,J10 | 1PPS | JUMPER |
| 12 | 1 | J11 | SMA 5 PIN | BNC-SIDE5 |
| 13 | 1 | LED1 | GREEN A | CC0603(M) |
| 14 | 1 | LED2 | GREEN B | CC0603(M) |
| 15 | 1 | LED3 | GREEN C | CC0603(M) |
| 16 | 1 | LED4 | BLUE A | CC0603(M) |
| 17 | 1 | LED5 | BLUE B | CC0603(M) |
| 18 | 1 | LED6 | YELLOW A | CC0603(M) |
| 19 | 1 | R21 | 10k | CC0402(M) |
| 20 | 2 | R22,R29 | 4.7K | CC0402(M) |
| 21 | 4 | R31,R47,R48,R49 | 330 | CC0402(M) |
| 22 | 1 | R36 | 2.2K | CC0402(M) |
| 23 | 4 | R37,R45,R50,R51 | 470 | CC0603(M) |
| 24 | 11 | R38,R39,R40,R41,R42,R43,R44,R52,R53,R54,R55 | 47 | CC0603(M) |
| 25 | 1 | R56 | 150 | CC0603(M) |
| 26 | 1 | R57 | 330 | CC0603(M) |
| 27 | 4 | R58,R59,R60,R61 | 100 | CC0603(M) |
| 28 | 2 | SW1,SW2 | EVQ-PAD04M | EVQ-PAD04M |
| 29 | 2 | U1,U2 | ESDA6V1BC6 | SOT-23-6L |
| 30 | 2 | U3,U4 | CP2102-GM | QFN28 |
| 31 | 1 | U5 | 74HCT373D | SO(20) |
| 32 | 1 | U6 | iTRAX03 | iTrax03 |
| 33 | 1 | U7 | ADP3339AKC-3.3 | SOT-223_1 |
| 34 | 1 | U8 | AAT3221IGV-2.8 | SOT25 |

# Appendix B – Board Design Details

## *VDDDIG regulator*

The ADP3339AKC-3.3[3] Regulator was selected as the main power supply for the iTRAX-03 receiver and for the VDDRF regulator.

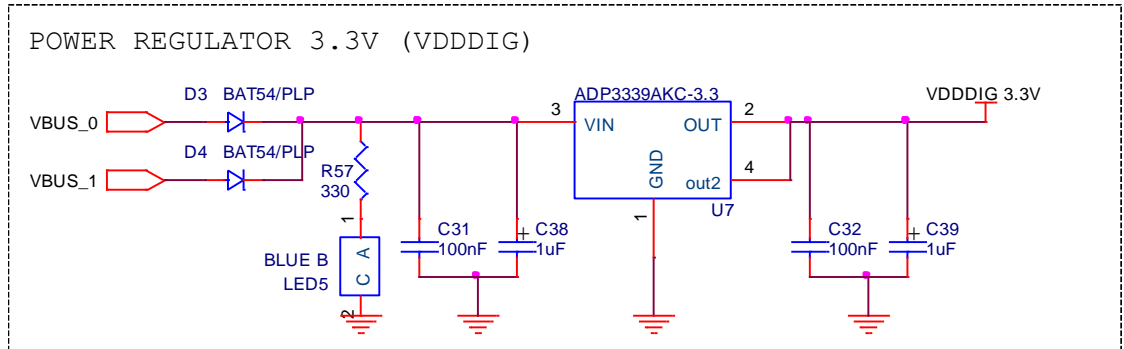

**Figure 12 - VDDDIG circuit**

A Led indicator was added to indicate proper power supply from one of the USB inputs.

## *VDDRF regulator*

The iTRAX-03 receiver RF modules are fed by a low ripple linear regulator. The regulator we selected is the AAT3221IGV-2.8-T1 150mA Low Dropout Linear Regulators by Analog Tech [2]. (One of the design constraints for the receiver was to limit VDDRF supply to 150mA). This device features extremely low quiescent current which is typically 1.1µA.

The AAT3221 type regulator has output short circuit and over current protection. In addition, the device also has an over temperature protection circuit, which will shutdown the LDO regulator during extended over-current events. We selected the 2.8V version of the AAT3221 regulator as recommended in [1].

In our application we've shortened the enable pin to the input and added a LED indicator for input voltage. The input is provided by VDDDIG.

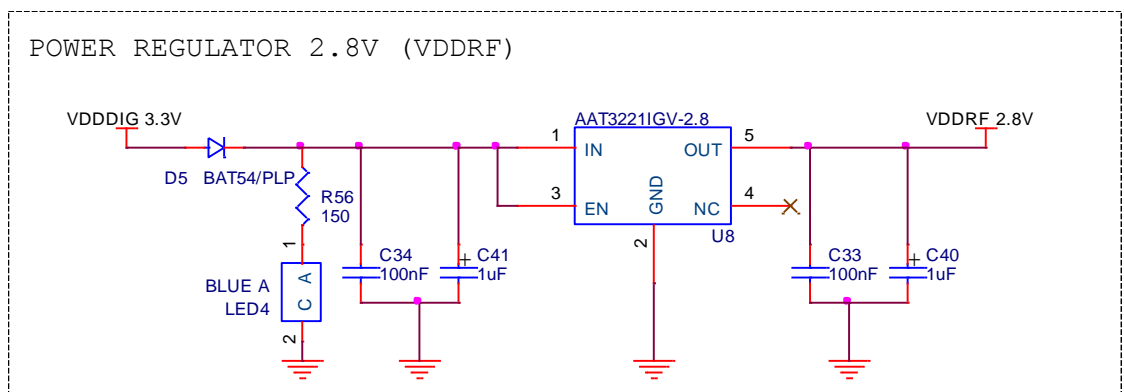

**Figure 13 - VDDRF circuit**

## *USB to UART bridge CP2102*

The CP2102[7] is a highly-integrated USB-to-UART Bridge Controller providing a simple solution for updating RS-232 designs to USB using a minimum of components and PCB space. The CP2102 includes a USB 2.0 full-speed function controller, USB transceiver, oscillator, EEPROM, and asynchronous serial data bus (UART) with full modem control signals in a compact 5 x 5 mm QFN-28 package.

In our design both CP2102 controllers are protected by ESD protectors. The controllers are fed by 5V USB supply and are connected to mini-USB connectors.

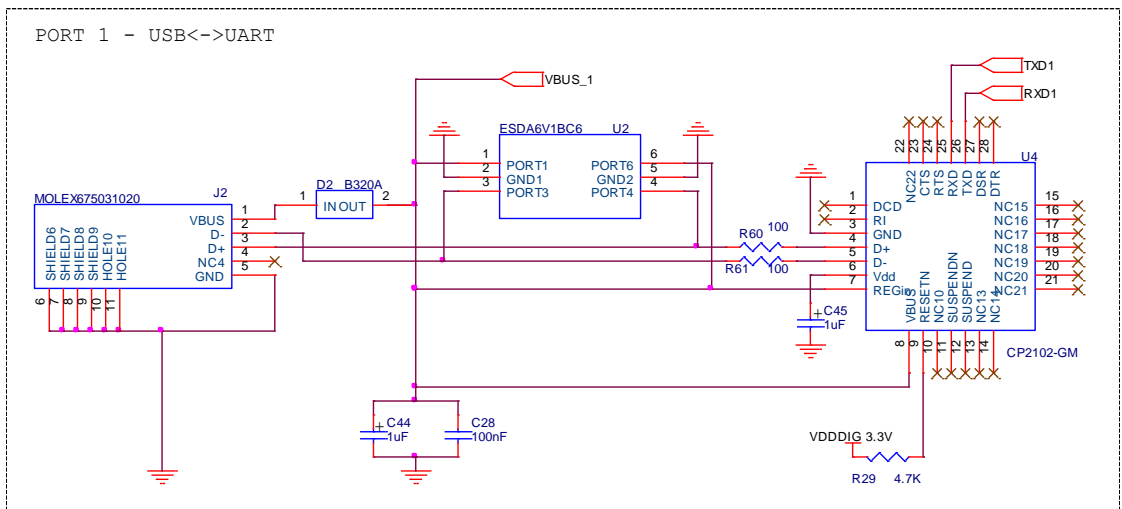

**Figure 14 - CP2102 USB to UART bridge**

## *iTrax03*

The core of the circuit is the iTrax03 receiver module. The receiver integrates the GPS chip set: RF down-converter uN8021 and uN8130 processor. The processor includes all base band functions needed for GPS signal acquisition, tracking and navigation. Dedicated high-performance search engine architecture enables a rapid search of visible satellites. 12 channel tracking unit insures that positioning is possible even in severe conditions.

The antenna input supports passive and active antennas and provides also an internal antenna bias supply.

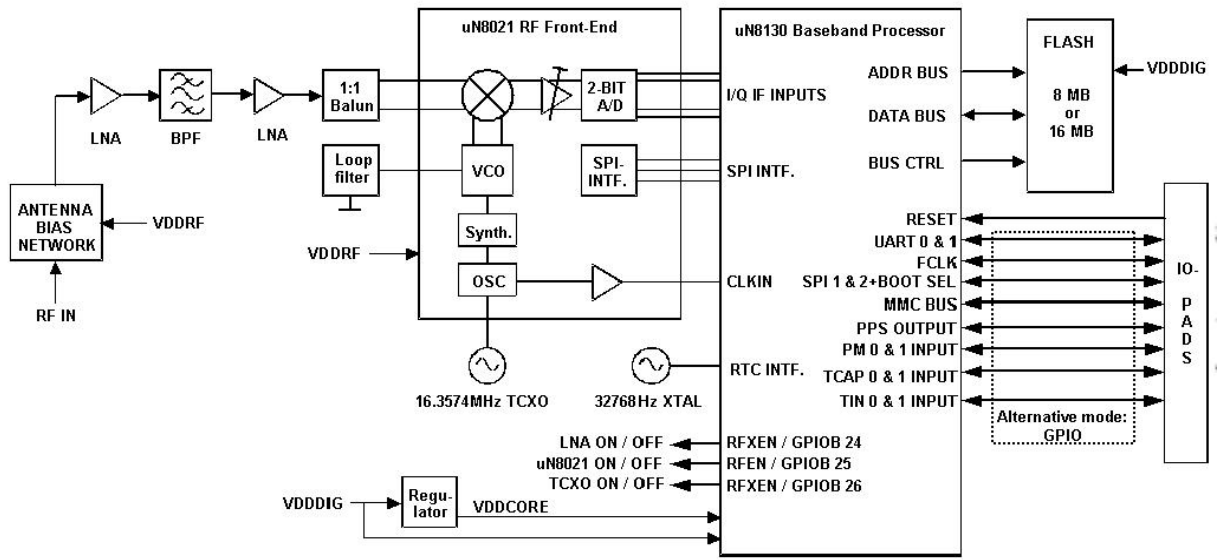

**Figure 15 - iTrax03 Block Diagram**

**iTrax03 Specifications:**

| | |
|---|---|
| Receiver | GPS L1 C/A code |
| Channels | 12 |
| Update rate | 1 Hz |
| Supply voltage (VDDRF) | +2.7…+3.3V, low ripple 2mV RMS |
| Supply voltage (VDDDIG) | +2.7…+3.3V |
| Antenna bias voltage | Same as VDDRF |
| Serial Port Configuration | Port 0 : NMEA (iTalk optional) Port 1 : iTalk |
| Serial data speed | NMEA: 4800 Baud, iTalk: 115200 Baud. |
| Power dissipation | 500mW max |
| Current output on antenna input | 0-100mA |

In our integration of the iTrax03 into the board, we've added reset and boot switches as described in the following circuit:

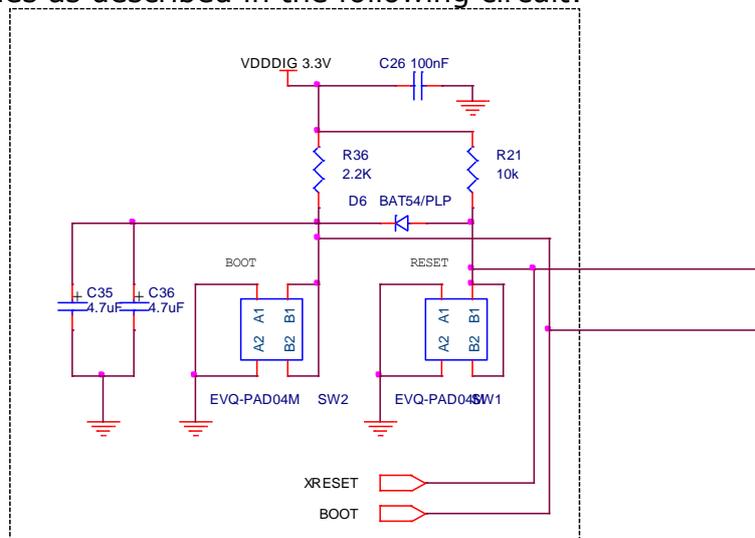

**Figure 16 - Reset & Boot Switches**

The receiver was connected to SMA connector for antenna input, VDDDIG and VDDRF inputs, RX and TX channels from the USB-UART converters, Reset and Boot switches and led indications:

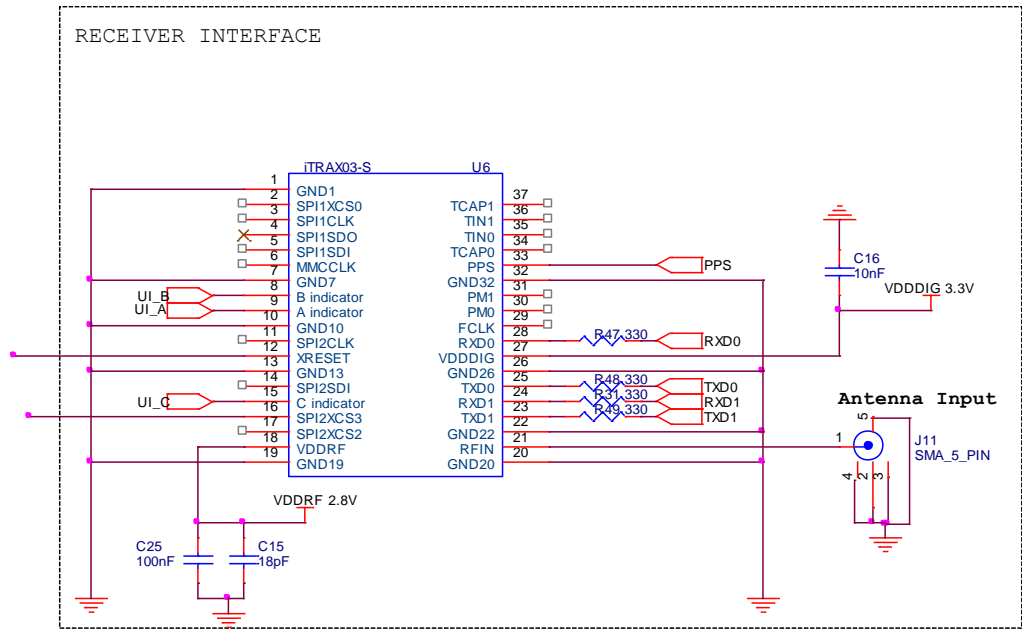

**Figure 17 - iTrax03 peripherals**

## 1PPS driver

The iTrax03-S receiver raises 1PPS output every 1 second, synchronized to the transfer of the data messages to the UART output. This signal may be used in future applications.

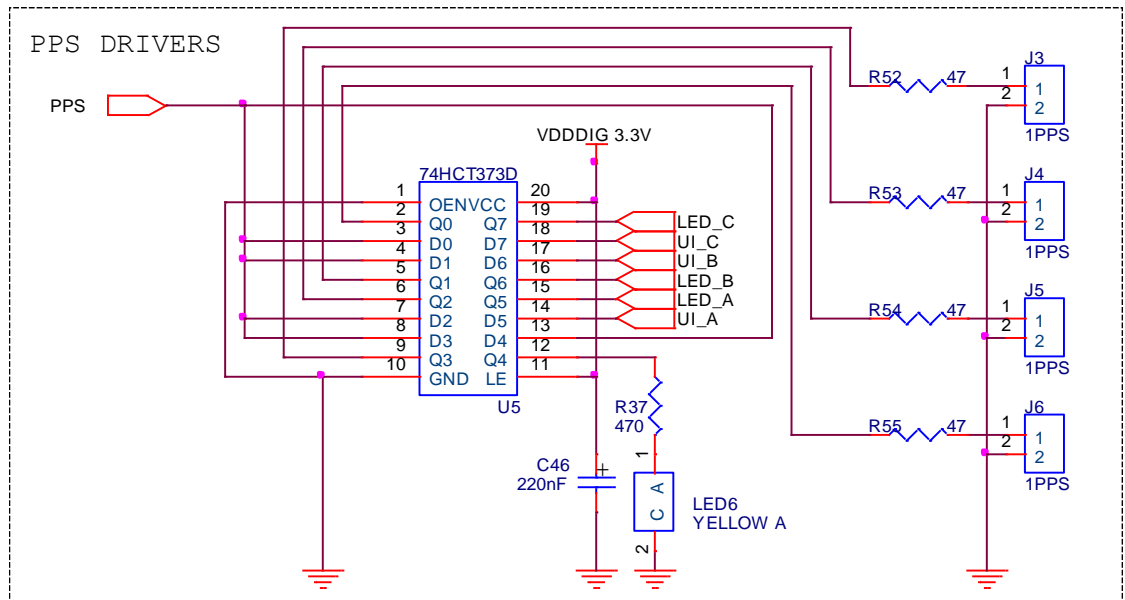

**Figure 18 – 1PPS driver**

Our implementation uses the 8 outputs of the 74HCT373D buffer in order to drive 4 1PPS outputs, 3 indication LEDs of the receiver operation status and an additional LED indicating proper output of the 1PPS.

## PCB layout

The PCB layout was designed by an external contractor and tested to meet the design. The **B**ill **O**f **M**aterials (BOM) is attached to this document (See Appendix A).

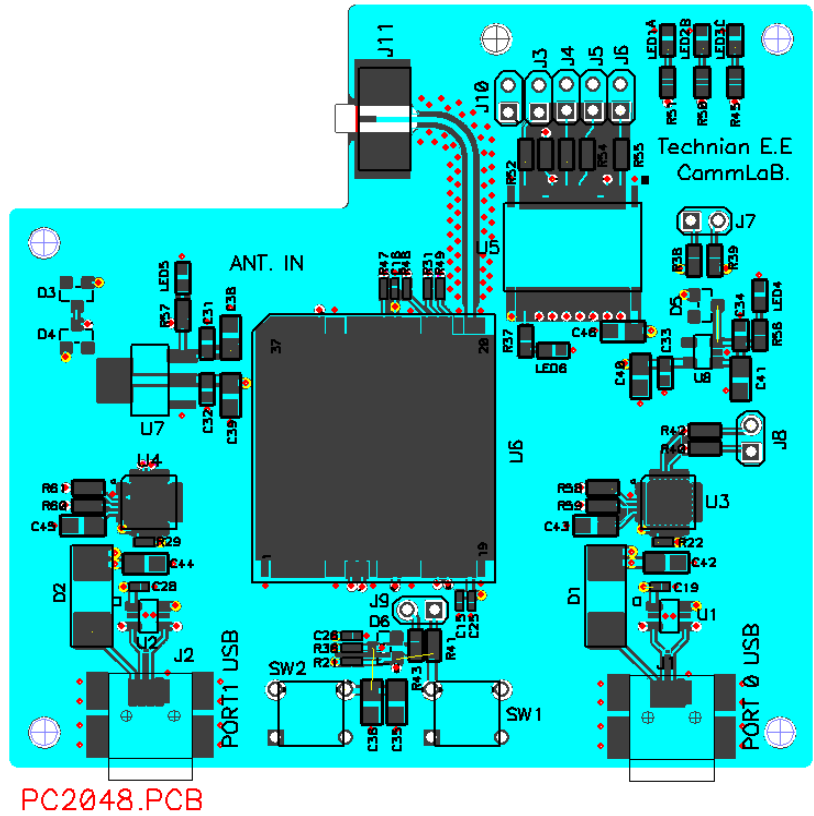

**Figure 19 – original PCB layout**

# Appendix C – RINEX protocol

One of the major problems with the iTrax03 receiver is the lack of standard RINEX[4] protocol output. The RINEX (**R**eceiver **In**dependent **Ex**change Format) protocol is used mainly in expensive geodesic receivers as it contains data irrelevant to most commercial receivers. Instead, the iTrax03 provides the output data in commercial NMEA protocol and also in advanced iTalk[5] binary protocol, suitable to more advances applications such as ours.

Therefore we were required to produce a Matlab code, responsible for translation of the relevant data from the iTalk messages to the relevant RINEX files. The MATLAB algorithm responsible for calculation of the DD (**D**ouble **D**ifference) is based on input of geodesic receivers with 2 types of files used: Navigation and Observation.

Each file type consists of a header section and a data section. The header section contains global information for the entire file and is placed at the beginning of the file.

## *RINEX structure*

This paragraph describes each data element required for the RINEX navigation and observation files, and its' source in the iTalk messages (Some of the values should be multiplied by a constant as described below):

## Navigation file

The header contains only the time of the conversion and name of the converting software. The body contains the following data, duplicated for each SV detected during the last update:

| RINEX | Source – iTalk | Description |
|---|---|---|
| SV number | (RawEph.wPRN) | |
| Year | (nav.UTC.wYear) | |
| Month | (nav.UTC.Month) | |
| Day | (nav.UTC.Day) | |
| Hour | (nav.UTC.Hour) | |
| Minute | (nav.UTC.Minute) | |
| Second | (nav.UTC) | |
| dAf0 | (RawEph.Clock.lAf0)* $2^{-31}$ | SV clock bias (seconds) |
| dAf1 | (RawEph.Clock.iAf1)* $2^{-43}$ | SV clock drift (sec/sec) |
| dAf2 | (RawEph.Clock.iAf2)* $2^{-55}$ | SV clock drift rate (sec/sec$^2$) |
| IODE | (RawEph.Orbit.wIODE) | IODE Issue of Data |
| Crs | (RawEph.Orbit.iCrs)* $2^{-5}$ | Amplitude of the sine correction term to the orbit radius (meters) |
| Delta n | Pi*(RawEph.Orbit.iDeltan)* $2^{-43}$ | Mean motion diff (radians/sec) |
| M0 | Pi*(RawEph.Orbit.IM0)* $2^{-31}$ | Mean anomaly at reference time (radians) |
| Cuc | (RawEph.Orbit.iCuc)* $2^{-29}$ | Amplitude of the cosine correction to the argument of latitude(radians) |
| e | (RawEph.Orbit.IEcc)* $2^{-33}$ | Eccentricity |
| Cus | (RawEph.Orbit.iCus)* $2^{-29}$ | Amplitude of the sine correction to the argument of latitude (radians) |
| sqrt(A) | (RawEph.Orbit.ISqrta)* $2^{-19}$ | Square of semi-major axis (m$^{0.5}$) |
| Toe | (RawEph.Orbit.dwToe) | Time of Ephemeris (sec GPS time) |
| Cic | (RawEph.Orbit.iCic)* $2^{-29}$ | Amplitude of the cosine correction to the angle of inclination (radians) |
| OMEGA | Pi*(RawEph.Orbit.IOmega0)* $2^{-31}$ | Long. of the ascending node of orbit plane at weekly epoch (radians) |
| CIS | (RawEph.Orbit.iCis)* $2^{-29}$ | Amplitude of the sine correction to the angle of inclination (radians) |
| i0 | Pi* (RawEph.Orbit.II0) * $2^{-31}$ | Inclination at ref time (radians) |
| Crc | RawEph.Orbit.iCrc * 2^-5 | Amplitude of the cosine correction to the orbit radius (meters) |
| omega | Pi*(RawEph.Orbit.IOmega)* $2^{-31}$ | Argument of perigee (radians) |
| OMEGA DOT | Pi*(RawEph.Orbit.IOmegaDot)* $2^{-43}$ | The derivation of $\Omega$ (radians/sec) |
| IDOT | Pi*(RawEph.Orbit.iIdot)* $2^{-43}$ | Inclination derivation (radians/sec) |
| Codes on L2 | 0 | N/A |
| GPS Week # | (RawEph.Clock.wWeek) | |
| Spare | 0 | N/A |
| L2 P flag | 0 | N/A |
| SV accuracy | 32 | [m] |
| SV health | (RawEph.wHealth) | |
| TGD | (RawEph.Clock.iGroupDelay)* $2^{-31}$ | Time group delay (seconds) |
| IODC | RawEph.Clock.wIODC | Issue of Data, Clock |
| Trans. time | RawEph.Clock.dwTowMs/1000 | Time of Week (ms) |
| Fit interval | RawEph.wFitPeriod | (hours) |
| Spare | 0 | N/A |
| Spare | 0 | N/A |

## Observation file

The header contains the following data:

| RINEX | Source – iTalk | Description |
|---|---|---|
| X | nav.WGS.IX (first epoch in file) | APPROX POSITION X (WGS84) |
| Y | nav.WGS.IY (first epoch in file) | APPROX POSITION Y (WGS84) |
| Z | nav.WGS.IZ (first epoch in file) | APPROX POSITION Z (WGS84) |
| Types of observation | N/A | C1 – Pseudo range from L1 <br> L1 - Carrier phase from L1 <br> D1 – Doppler frequency from L1 |
| Time of first observation | nav.UTC (first epoch) | |
| Time of last observation | nav.UTC (last epoch) | |

The Body contains the following data for each epoch:

| RINEX | Source – iTalk | Description |
|---|---|---|
| SV number | (RawEph.wPRN) | |
| Year | (nav.UTC.wYear) | |
| Month | (nav.UTC.Month) | |
| Day | (nav.UTC.Day) | |
| Hour | (nav.UTC.Hour) | |
| Minute | (nav.UTC.Minute) | |
| Second | (nav.UTC) | |
| Active SV list | track.swLockedChInfo (8 low bits) | List of active GPS SVs, preceded by 'G' symbol. |
| Pseudo Range | pseudo.dPseudoRange | PR = distance + c * (receiver clock offset - satellite clock offset + other biases) (meters) |
| Carrier phase | track.dwCarrCount(ep,idx) + bitand(abs(track.wCarrPhase(ep,idx)), 255)/256.0 | The phase is the carrier-phase measured in whole cycles |
| Doppler Frequency | pseudo.dDoppler | (Hz) |

# Appendix D - iTalk to RINEX conversion

## *MATLAB GUI*

We've created user-friendly interface for converting the iTalk archive files to standard RINEX files and processing the RINEX files. The following options are available through our GUI interface:

1. Conversion mode:
   a. Manual Select – User is given the option to select one of the calculated sequences (# of SVs and Number of epochs).
   b. All – All calculated sequences will be converted to multiple RINEX files.
2. Print Options (available only in manual conversion mode):
   a. SV vs. Epoch graphs are displayed for selected sequence.
   b. Elevation vs. Epoch graphs are displayed for selected sequence.
3. Parameters:
   a. Minimum SV elevation to process (as calculated from iTalk archive files)
   b. Baseline length in meters.
4. Working directory – source of the iTalk files and destination of the RINEX directory.

The user can select whether to only convert the iTalk files to RINEX or to also determinate the baseline using the DD calculation algorithm [6]. (The latter is available in 'Manual Select' mode).

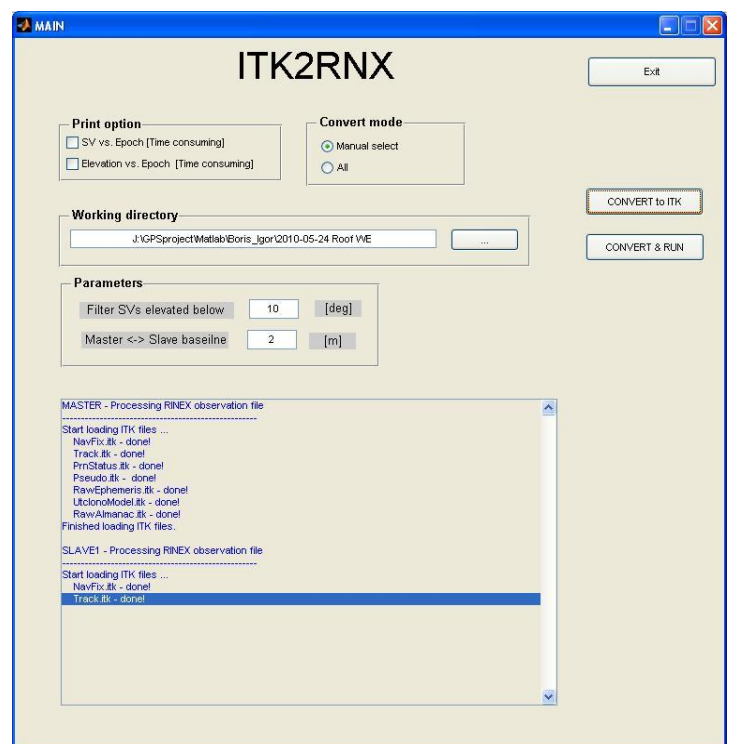

**Figure 20 – ITK2RNX GUI**

### *Reading iTalk files*

Our next mission was to load the data elements (described in chapter 6) into convenient database that contains all relevant data from the receivers. This was accomplished using **ITK2RNX_LoadITK.m**. This Matlab file utilizes loading relevant data from .itk binary files into matrices containing relevant data. The file uses the following functions provided by Fastrax as part of the SDK (some of the functions were modified from the original in order to minimize processing time):

| File name | Contains |
|---|---|
| read03nav.m | Navigation data |
| read03track.m | Track process data |
| read03prnstatus.m | PRN (SV) status data |
| read03pseudo.m | Pseudo process data |
| read03RawEph.m | Raw Ephemeris data |
| read03ionotropo.m | Ionotropic information |
| read03rawalm.m | Raw almanac data |

### *Synchronization of ITK data*

**Problem:** We've discovered that the dwTick field (responsible for synchronization of the data in different files) in pseudo.itk and track.itk files is not always synchronized. For some of the readings, a shift of ±1 tick (1 sec) is found.
**Solution: ITK2RNX_AlignTick.m** solves this problem by taking the common parts of the files.

### *Processing SV data*

The iTalk protocol provides the SV information in processing channels. Each receiver has 12 channels, containing the SV data. In order to process data from different receivers, we had to transfer the data from 'channel' to 'SV number' axis. This process is performed by **ITK2RNX_SVstats.m**. This functions returns the following data elements:
1. **Data** - containing SV availability along epochs
2. **Elev -** containing SV elevation along epochs
3. **TimeStamp** – Start and end time values in GPS Time of Week ms format.
4. **NonValVec** – Vector containing the epochs declared as non-valid by the receiver.
5. **JumpVec** – Vector containing the epochs where time jumps are found (After neglecting the non-valid epochs).

### *Mutual time and elevation synchronization*

The measurements from the 2 receivers can be taken in different time periods. Therefore we need to select only the mutual time periods according to the timestamps calculated in SVstats. In addition the data is filtered by the minimum elevation criteria, as defined by user in the GUI.

Due to atmospheric conditions and various other environmental influences (physical obstacles, clouds and so on), different SVs may be tracked by the 2 receivers at the same time. Our algorithm generates one data structure (OverlapAll2) which contains the common SVs logical AND in order to find overlapping sequences.

## *Calculation of longest sequence*

In order to achieve the best results from the DD calculation algorithm [6] ,we must find the longest non-interrupted sequence for each number of SVs above 4. This action is performed for each epoch by **ITK2RNX_SV_filter_perm.m**. This function goes epoch by epoch, takes list of SVs available at this epoch and searches the longest uninterrupted sequence for all available permutations of this SVs.

For example:  if SVs 1,2,3,4,5,6 are available the following permutations are checked for longest sequence:
[1,2,3,4,5], [2,3,4,5,6], [1,3,4,5,6], [1,2,4,5,6], [1,2,3,4,6], [1,2,3,4,6], [1,2,3,4,5,6]

## *Generation of RINEX files*

Our code generates 2 types of RINEX files as described in chapter 6: Observation and Navigation.

The navigation file is generated mainly using data received from *RawEph.itk* file, after correction by constants.

The observation files (one for each receiver) are generated mainly using data received from *pseudo.itk* and *track.itk* files after being processed as described in the paragraphs above. The main data used for baseline determination is the carrier phase calculated by the following formula:

Φ(cycles) = dwCarrCount + wCarrPhase / 256 [8]